# GREI Data Repository AI Taxonomy

*Authors: John Chodacki (California Digital Library), Mark Hanhel (figshare), Stefano Iacus (Dataverse), Ryan Scherle (Dryad), Eric Olson (Center for Open Science), Nici Pfeiffer (Center for Open Science), Kristi Holmes (Zenodo), Mohammad Hosseini (Zenodo)*

The [Generalist Repository Ecosystem Initiative (GREI)](#) is a NIH-funded program where repositories collaborate in a "coopetition" model to enhance the work of generalist data repositories, which are critical infrastructure across research domains. As part of our commitment to this work, we recognize the evolving importance of artificial intelligence (AI) in the future of science and infrastructure. To help our community navigate AI-driven changes, we have developed the following taxonomy to illustrate the roles AI can play in managing data repositories, improving data quality, and increasing accessibility.

Building on previously developed taxonomies and our coopetition efforts, the GREI repositories propose the following "GREI Data Repository AI Taxonomy," specifically tailored for data repository roles.

## Why do we need this?

A taxonomy like the one outlined below holds significant importance for GREI and other domain-specific or institutional data repositories, serving as a structured framework to develop strategies, facilitate discussions, and effectively address the opportunities and challenges presented by AI integration.

Firstly, this taxonomy can offer a systematic classification of the distinct roles AI can play in data repository operations. This classification enables a comprehensive understanding of AI's potential contributions to repository operations, which, in turn, assists in strategic planning. By identifying specific areas where AI can bring the most value, interested parties can allocate resources efficiently, aligning AI initiatives with their organizational objectives.

Secondly, the taxonomy provides a common language and reference point, fostering productive discussions among participants. It ensures that organizations can communicate effectively about their AI workflow integrations, share best practices, and collaboratively address challenges. This shared understanding of AI's role in data management promotes innovation and helps ensure that AI initiatives adhere to ethical and quality standards.

Additionally, the taxonomy can serve as a structured framework for addressing challenges related to AI integration in data repositories. By categorizing AI involvement into distinct areas, potential issues and ethical concerns can be systematically identified and explored . This structured approach enables the establishment of safeguards, transparent AI practices, and the incorporation of human oversight where necessary. It also helps in building trust with users, addressing concerns related to data accuracy, transparency, security and integrity.

# The GREI Data Repository AI Taxonomy

Just as AI can revolutionize other forms of scholarly communications like peer-reviewed publications ([Ref](#)), it can bring significant improvements to data repositories ([Ref](#)). This taxonomy outlines seven areas for AI in data repository management:

1. **Acquire**: Efficiently gather, collect, and ingest data and metadata from various sources, including researchers, sensors, and external datasets.
2. **Validate**: Ensure the quality, accuracy, and integrity of the data and metadata to maintain its reliability and trustworthiness.
3. **Organize**: Categorize, structure, and catalog data and metadata to facilitate easy retrieval, analysis, and sharing.
4. **Enhance**: Enrich and augment data and metadata with metadata, annotations, or standardized formats to improve its utility and interoperability.
5. **Analyze**: Employ AI-driven analytics to uncover insights, patterns, and trends within the data and metadata, aiding researchers and decision-makers.
6. **Share**: Facilitate the discovery, access, and distribution of data and metadata within and beyond the repository, promoting collaboration and knowledge dissemination.
7. **Support:** Provide suggestions and answer questions for users of the data and metadata, including the users who are submitting, the users who are consuming, and the repository staff.

This taxonomy aims to provide a framework for understanding the specific roles that AI can play in data management and specifically data repository work including data discovery, documentation, and reuse. To further illustrate each category, we present these examples:

## Acquire

The category of "Acquire" is important because efficiently gathering and ingesting data from researchers and other sources is the foundational step in data management. AI can streamline this process by automating data collection, saving time and reducing the risk of human errors. The ability to acquire data effectively ensures that repositories can stay up-to-date with the latest research and maintain a comprehensive database. Moreover, AI can be employed to adapt to the growing volume of data sources, making it crucial for categorizing AI usage in data repository work.

Examples of work in this area could include:
- AI-assisted ingestion of research data from online sources.
- Integrate data from various instruments and devices into the repository.
- Aggregate data submissions from researchers and collaborators.

## Validate

Data quality and reliability are non-negotiable in data repositories. "Validate" is a critical category because AI can play a pivotal role in verifying data accuracy and authenticity. Inaccurate or unreliable data can lead to erroneous research findings, so AI tools can cross-reference data against established standards, identify anomalies, and validate the source of incoming data. Proper validation ensures that researchers can trust the data they access, making this category indispensable in the taxonomy.

As AI evolves, examples of work in this area could include:
- AI-assisted identification and flagging of potential data anomalies or errors.
- Cross-reference data against established standards and guidelines.
- Verify the authenticity and source of incoming data.

## Organize

Effective data organization and description is key to making data repositories accessible and usable. AI can categorize, structure, and catalog data efficiently, ensuring that data is stored in a way that facilitates easy retrieval and analysis. Without proper organization, data repositories become unwieldy and less valuable to researchers. This category acknowledges the central role AI plays in ensuring that data is structured and categorized for maximum utility.

Examples of work in this area could include:
- Categorize data by research discipline, project, or data type.
- Create hierarchical structures to accommodate complex datasets.
- Tag data with relevant keywords and descriptors for easy search.

## Enhance

Enriching data with metadata, annotations, or standardized formats is crucial for ensuring data interoperability and usefulness. AI can significantly contribute by generating descriptive metadata or converting data to standardized schemas. This category underscores AI's role in enhancing data quality and making it more accessible to a broader audience. Nevertheless, it also emphasizes the need for human oversight to ensure that enhancements align with the data's intended purpose.

Examples of work in this area could include:
- AI-assisted generation of metadata and descriptions for data entries.
- Convert data formats to standardized schemas or ontologies.
- Translate metadata and content to multiple languages for global accessibility.
- Build knowledge graphs across ontologies that might assist in new discoveries

## Analyze

The "Analyze" category highlights AI's capability to extract insights, patterns, and trends from the vast amount of data within repositories. AI-driven data analytics tools can provide valuable insights that may not be immediately evident to human curators. These insights can guide research and decision-making processes. However, it's equally important to categorize this area as it reminds us that human interpretation is essential to ensure that AI-generated findings are contextually accurate and meaningful.

Examples of work in this area could include:
- Utilize AI-driven algorithms to identify correlations and trends in large datasets.
- Perform sentiment analysis on textual data within the repository.
- Assist in describing images for accessibility and other use cases.
- Suggest datasets that are similar to a given dataset, aiding researchers in finding relevant information.
- Create video to text transcriptions using AI algorithms

## Share

The "Share" category is pivotal because data repositories exist to disseminate knowledge and facilitate reuse. AI can assist in recommending relevant datasets, securing data access, and generating citations. It is essential to categorize AI's role in data sharing as it not only improves the efficiency of data distribution but also underlines the need for human-driven decisions regarding access controls and the suitability of sharing mechanisms. Trust and transparency in data sharing processes are paramount, and AI can support these aspects effectively.

Examples of work in this area could include:
- Recommend relevant datasets to researchers based on their interests and previous usage.
- Facilitate data sharing and collaboration through secure access controls and sharing options.
- AI-assisted generation of citations and references for data used in research publications.

## Support

The "Support" category is crucial for ensuring that users, whether they are data submitters, data consumers, or repository staff, can effectively interact with and benefit from the data repository. AI in this context acts as a bridge, providing intelligent assistance, automating routine queries, and facilitating a smoother user experience. This category recognizes that AI can enhance the overall user support ecosystem, making it more responsive, personalized, and efficient.

Examples of work in this area could include:
- Provide immediate, intelligent support to users with AI-driven chatbots and virtual assistants.
- Analyze user behavior to offer personalized recommendations, improving access to relevant datasets and resources.
- Update documentation and tutorials, ensuring users have access to latest guidance.
- Real-time feedback on data submissions, helping maintain high data quality standards.

- Enhance search functionality by understanding natural language queries and providing accurate, contextual suggestions.

## Balancing AI and Human Expertise in Data Repositories

The integration of AI in data repositories offers significant opportunities for enhancing efficiency, data quality, and user experience. However, the successful implementation of AI requires a careful balance between automated processes and human expertise. While AI can handle large-scale data processing tasks with speed and precision, human oversight remains crucial for ensuring that these processes align with ethical standards and meet the nuanced demands of research communities.

AI's role in data repositories spans from data acquisition and validation to organization and sharing, as outlined in the above GREI Data Repository AI Taxonomy. However, at every stage, the balance between AI-driven automation and human intervention must be meticulously managed. For example, while AI could potentially ingest and validate vast amounts of data, human curators must also verify that AI-generated outputs adhere to the necessary quality standards and are contextually accurate. This human oversight ensures that the data remains reliable and trustworthy, particularly in fields where even minor errors can have significant consequences.

One effective approach to maintaining this balance is the implementation of tiered AI automation levels, where the degree of automation is matched to the task's complexity and the potential impact of errors. For example:

- **No Automation**: Human experts handle tasks entirely, appropriate for highly specialized or sensitive data.
- **Automation with 100% Certainty**: AI handles tasks where outcomes are highly predictable, such as formatting metadata to standardized schemas.
- **AI with Human Review of Each Change**: AI suggests actions that require human approval, ensuring that any AI-driven modifications meet the required standards.
- **AI with Human Review of a Sample of Changes**: In less critical tasks, humans might review only a sample of AI-driven changes to ensure quality while benefiting from the efficiency of automation.
- **Fully Autonomous AI**: For routine, low-risk tasks, AI operates independently, freeing human experts to focus on more complex issues.

This structured approach allows repositories to maximize the benefits of AI while safeguarding data integrity and trust. The transparency of AI processes is critical. Repositories must clearly communicate to users when and how AI is being used, especially in tasks that directly impact data quality or accessibility. Providing users with this transparency builds trust and allows for informed decision-making, as users can understand the role AI plays in the data they are accessing or contributing.

The provenance of any changes or enhancements created by non-human systems, such as AI, is critical for ensuring data integrity, trustworthiness, and transparency in data repositories. Without clear provenance, it becomes difficult to verify the origins of data, understand the rationale behind AI-driven changes, or correct any potential errors introduced by automation. Repository systems should endeavor to expose this provenance in the GUI of the repository and metadata of any API retrieved records.

## Trust and Transparency in Data Management

As AI becomes more integrated into data repository workflows, trust remains a crucial factor. To build and maintain trust, it's essential to adhere to regulatory frameworks such as [the Framework for Federal Scientific Integrity Policy and Practice](#) developed by the Office of Science and Technology Policy (OSTP). For generalist repositories and their users to adhere to regulatory frameworks, ethical behavior that demonstrates honesty, transparency, and trust should be consistently practiced. Towards this end, three challenges are noticeable. First, generalist repositories strive to ensure that their infrastructure including the interface, instructions on how to use AI are up to date. However, given the range of international ethical and regulatory frameworks and their differences (e.g., [AI Act](#) in Europe versus [AI Executive Order](#) in the US), adhering to one regulatory framework without flouting/trespassing other frameworks might be challenging. Second, generalist repositories regularly prompt, and if necessary, apply pressure (e.g., screening and quality control) to ensure that uploaded content meets certain standards. Nevertheless, the sheer amount and diversity of newly uploaded content, and the range of roles conducted by AI combined with different levels of information/digital literacy, challenges adherence to regulatory frameworks. Third, since content is openly available and generalist repositories cannot control reuse and AI digestion of their data, they may inadvertently facilitate misuse or harm to individuals and groups, thereby eroding the public trust in science.

We propose three suggestions to promote trust and transparency when using AI in generalist repositories. 1) Devising balanced AI governance and stewardship policies, considerate of the diverse disciplinary and international ethical/legal norms that may affect AI use and potential data transfers (e.g., GDPR versus HIPAA); 2) Developing codes of practice related to AI use and signposting for users (e.g., tag AI-generated/-manipulated content, identify AI's extent of involvement and limitations); 3) Initiating work to anticipate how content could be used (and misused) by various AI applications to provide adequate disclaimers (and if needed, restrictions).

## Conclusion

AI has the potential to revolutionize data repository management, improving efficiency, data quality, and accessibility. By adapting this GREI Data Repository AI Taxonomy and embracing a balanced approach that combines AI and human expertise, the wider data repository community can harness the full potential of AI while maintaining data integrity and ethical standards. Each of these taxonomic areas is uniquely important in categorizing the usage of AI in data repository work because they represent distinct stages in the data management process, and AI's contribution to each stage is pivotal in achieving efficient, reliable, and accessible data repositories. Categorizing these roles ensures that the specific strengths and responsibilities of AI within data management are acknowledged and properly managed.

As we move forward, we must continue to explore and iterate on the application of AI in data management, building a strong foundation for the future of research and knowledge sharing across GREI repositories as well as the wider repository community.

——

Many thanks to the authoring team of "[An Initial Scholarly AI Taxonomy](#)" which offered inspiration for this article. In addition, thank you to NIH and the [Generalist Repository Ecosystem Initiative (GREI)](#) for supporting us writing this. This work was supported by the National Institutes of Health (NIH) Office of Data Science Strategy/Office of the NIH Director pursuant to OTA-21-009, "Generalist Repository Ecosystem Initiative (GREI)" through Other Transactions Agreement (OTA) Numbers OT2DB000001, OT2DB000002, OT2DB000003, OT2DB000004, OT2DB000005, OT2DB000006, and OT2DB000013.